# Comparison of SNSPDs biased with microwave and direct currents

S. Doerner, A. Kuzmin, S. Wuensch, Member, IEEE, and M. Siegel

*Abstract* —This paper presents a detailed investigation of superconducting nanowire single-photon detectors (SNSPDs) biased with microwave and direct currents. We developed a hybrid detector, which allows the operation in the rf and dc operation mode. With this hybrid detector, we are able to compare the count rates of the same nanowire biased with dc and rf currents. Furthermore, we demonstrate the use of the oscillating current in the rf operation mode as a reference signal in a synchronous single-photon detection mode.

## I. Introduction

Superconducting nanowire single-photon detectors (SNSPDs) offer high detection efficiencies over a wide spectral range from visible- up to mid-infrared wavelength. They show very high count rates with only few dark-count events and a timing resolution of detected single-photons in the low ps range [1],[2]. These properties make SNSPDs a promising candidate for applications like space-to-ground laser communications [3], Quantum key distribution [4] or Fluorescence-lifetime imaging [5]. However, in order for the detectors to be used in such applications, multi-detector systems are needed. Those systems require a multiplexing scheme to reduce the number of wires between the cryogenic detector ambient and the readout electronics. Otherwise, the resulting heat load as well as the increasing complexity due to the high number of readout channels, would limit the number of pixels.

As a solution, we proposed in [6] the usage of a radio-frequency superconducting nanowire single-photon detector (RF-SNSPD). The idea of the RF-SNSPD is based on the embedding of a single-photon-sensitive nanowire in a lumped-element superconducting resonator. The resonator acts as a filter and defines the bandwidth of the single pixel in a frequency-division multiplexing scheme similar to microwave kinetic inductance detectors (MKIDs) [7]. In contrast to MKIDs the RF-SNSPD is operated very close to the critical current of the nanowire, using the oscillating microwave current inside the resonator. Thus the detector switches between the superconducting and normal conducting domain which creates a significant change of the quality factor due to single-photon absorption. Thus, we demonstrated the operation of a 16-pixel RF-SNPSD array with only one common feed line to connect the room temperature electronics with the cryogenic detector ambient [8].

In order to further increase the pixel number, to more than hundreds or even kilo-pixel arrays, a deeper understanding of the detector operation with microwave currents at high frequencies is necessary. So, in this paper we present a comparison between the achieved detection efficiencies of a hybrid detector operated with dc and rf bias. Furthermore, we demonstrate the gating of the detector by the oscillating microwave current at twice its resonant frequency $f_{res}$.

## II. Design of a hybrid detector

The achieved detection efficiencies of SNSPDs biased with dc and rf currents have already been compared in [6]. At that time, we presented the results of different detectors. However, to get a more accurate comparison, the results of the same nanowire biased with dc and rf currents must be compared. Therefore, we developed a new RF-SNSPD circuit which can be operated in both operation modes.

The design and the equivalent circuit scheme are shown in Fig. 1. The detector consists of a single-photon sensitive nanowire which is modeled as an inductor $L_{kin}$ and a resistor R. The resistor represents the normal conducting domain in the nanowire resulting from the absorption of a single photon. In

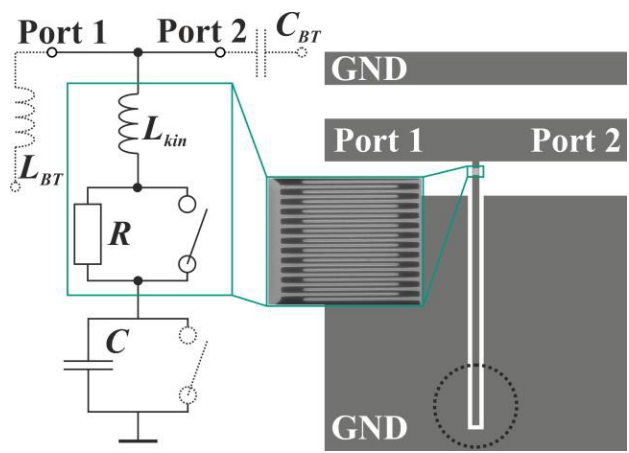

Fig. 1  Layout and equivalent circuit (ESB) of the hybrid detector. To change between rf and dc operation, a bond wire shorts the capacitor C, which is indicated with a dashed circle in the layout and by a dashed switch in the ESB. A further inductor $L_{BT}$ and capacitor $C_{BT}$ are also connected to port 1 and 2 to split the bias and readout signals.

This work was supported in part by the Karlsruhe School of Optics and Photonics (KSOP).

S. Doerner, A. Kuzmin, S. Wuensch and M. Siegel are with the Institute of Micro- and Nanoelectronic Systems (IMS), Karlsruhe Institute of Technology (KIT), 76187 Karlsruhe, Germany, (steffen.doerner@kit.edu).

the ready state, R is shorted by a parallel switch. Thus, the nanowire acts as the inductive part of a resonant circuit which is completed by a serial connected capacitor C. This resonant circuit is placed in the gap between the conductor and the ground plane of a coplanar waveguide (CPW) which is matched to the 50 Ω readout electronics. The nanowire is meander shaped over a length of 100 µm with a width of 100 nm. It is connected to the conductor on one side and to the capacitor C on the other. This capacitor is designed as an interdigital capacitor with only one large finger, extending into the ground plane over 350 µm. Hence, it links the resonant circuit with the ground plane of the CPW.

The entire circuit, consisting of the waveguide and the resonant circuit, is made of a 5 nm thick NbN film deposited on an R-plane sapphire substrate. In [8] the individual steps involved in the fabrication process can be found. The resulting values for $L_{kin}$ and C are calculated to 37.4 nH and 119 fF respectively using the sonnet em software [9]. Thus, the resonance frequency of the circuit is expected at 2.4 GHz.

In this configuration as a resonant circuit, the capacitor galvanically decouples the nanowire from the ground plane. To enable a detector operation with dc bias currents, the capacitor needs to be shunted. Therefore, we connected the capacitive finger and the ground plane using an indium bond wire, marked as a dashed circle in Fig. 1. In the equivalent circuit, the capacitor is now shorted by a switch, shown also in dashed lines. By connecting the inductor $L_{BT}$ and a capacitor $C_{BT}$ as a bias tee, the nanowire is now in a conventional dc configuration. Thus, we are able to characterize the detector in the rf operation mode first. After shorting the capacitor, we can then measure the optical detector properties again with the operating point set up by direct currents.

## III. MEASUREMENT AND RESULTS

For the measurements, the fabricated device is mounted into a gold coated brass housing which allows the connection of dc lines as well as coaxial cables. The housing is placed into a dip-stick cryostat and cooled down to the boiling temperature of liquid helium at 4.2 K. Inside the dip stick, there is also a multi-mode optical fiber which enables the illumination of the active area of the nanowire during the experiment. For the rf measurements, we placed cold attenuators (-32 dB) inside the rf lines as well as a cold microwave amplifier with a gain of 39 dB in the return path.

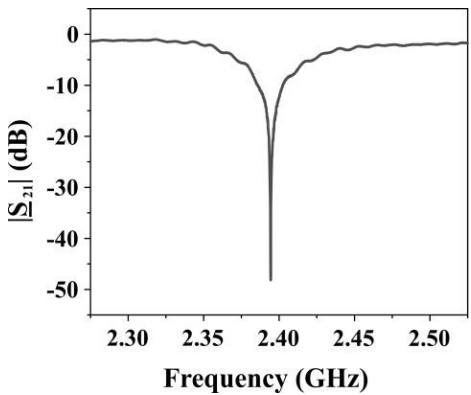

Fig. 2    Microwave transmission measurement of the coplanar waveguide. It shows the amplitude of the signal versus frequency. At 2.39 GHz the transmission is disturbed by the resonance of the RF-SNSPD.

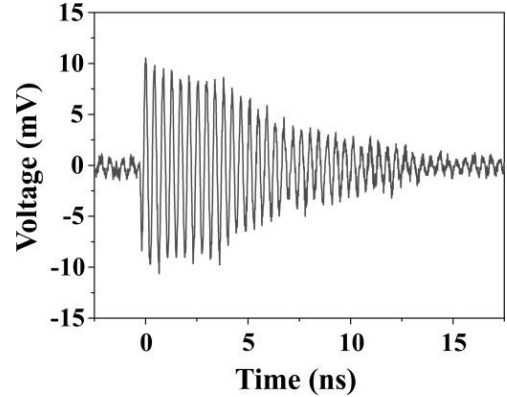

Fig. 3    Detector response with rf bias. It shows the measured voltage at port 2 over time. At t = 0 a response of the detector in the rf operation mode is visible.

To split the bias current and the detector response in the dc measurements a bias tee is needed. Therefore, an additional 680 nH inductor was connected to port 1 of the chip as well as a 12 nF capacitor to port 2.

### A.  Detector operation with rf bias

As a first step, we characterized the microwave properties of the device in a transmission measurement of the coplanar waveguide, using a vector network analyzer (VNA). Figure 2 shows the amplitudes of the transmitted signals from port 1 to port 2 at different frequencies. At 2.39 GHz a strong resonance is visible, which indicates the operating frequency of the detector in the rf mode. The loaded quality factor Q reaches 39 and is limited by the strong coupling of the resonant circuit to the external 50 Ω network.

Since the timing resolution of the VNA is not suitable to detect the fast amplitude changes of the carrier frequency due to single-photon absorption, we changed our setup. An analog signal generator is used to couple the bias tone at $f_{res}$ into the coplanar waveguide. The transmitted signal at port 2 is now measured by a real-time oscilloscope and analyzed by computer software. A recorded detector response after an absorbed single photon in this configuration is depicted in Fig. 3. If the detector is in the superconducting state its resistance is very small in the range of a few mΩ. Hence, the a strong mismatch between the resonant circuit and the 50 Ω coplanar waveguide, causes the main part of the applied bias power at port 1 to be reflected. Thus, the transmitted signal amplitude is damped by more than 50 dB and only noise is recorded by the oscilloscope.

At t = 0 a strong increase of the amplitude is visible due to the fast switching of the resistance inside of the nanowire. After the absorption of a single photon, the resulted normal conducting domain inside the nanowire of a few hundred Ω [10], damps the resonance completely. Thus, for a short time,

## C. Comparison of both operation modes

In both operation modes, we measured the count rates of the detector at different operation points and at different wavelengths of the incident photons onto the nanowire. From those results we calculated the detection efficiency for each wavelength as the ratio of the measured count rate to the number of incident photons onto the area of the nanowire. Therefore we measured the intensity of the fiber tip spot using a power meter and the beam distribution using a CCD camera, which was aligned in the same distance as the detector. The results are shown in Fig. 5.

For a direct comparison of the operating points, we converted the power levels in the rf operation mode to equivalent current amplitudes. In a first step we measure the critical output power of the generator, which drives the nanowire from the superconducting to the resistive state $P_{out,c}$. We assume this power level corresponds to the critical current $I_c$ measured in the dc operation mode. Further we use the scaling $i = C\sqrt{P_{Out}}$, where $C = I_C/\sqrt{P_{out,c}}$. In doing so, we are able to normalize any output power of the generator to equivalent current amplitudes inside the nanowire.

The comparison of both operation modes (Fig. 5) reveals similar count rates and thus, equivalent detection efficiencies if biased with rf or dc currents up to a level of 90 percent of the critical current. However, due to the oscillating operation point using rf bias currents the comparison is not quite fair.

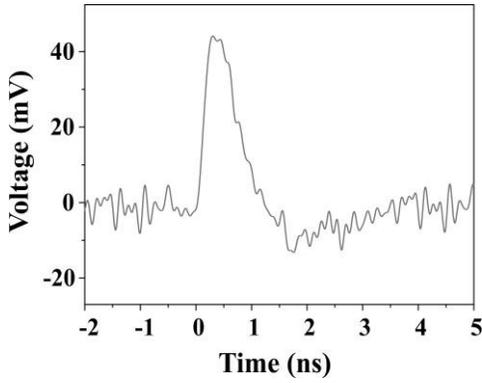

Fig. 4   Detector response operating the detector with dc bias. It shows the measured voltage response at port 2 over time. At t = 0 the response of the detector in the dc operation mode is visible.

the bias tone of the generator is transmitted undisturbed on the CPW. When the nanowire recovered the superconducting state, the resonant circuit starts with its transient oscillation defined by the quality factor Q. Hence, the decay time of the detector is calculated as $\tau = Q/(\pi f_{res}) \sim 5.2$ ns.

## B. Detector operation with dc bias

By shorting the capacitor C on the chip, the nanowire can be operated as a conventional dc biased SNSPD. Instead of the signal generator, a battery-powered dc source is connected in series to the inductor $L_{BT}$ at port 1. In this configuration we determined the critical current of the nanowire by 42.8 µA at 4.2 K. The detector response is still measured with the real time oscilloscope at port 2. The corresponding detector answer to an absorbed photon is shown in Fig. 4. At t = 0 ns the resistance of the nanowire starts to increase which causes a fast-rising voltage across the capacitor $C_{BT}$, measured as converted voltage signal at the 50 Ω input of the readout electronics. The amplitude level is higher now, because we changed the amplifier to a cascaded room temperature amplifier with gain of 69 dB. The decay time is now defined by $L_{kin}$ and the resistance of the readout channel $R_L$ as $\tau = L_{kin}/R_L \sim 0.75$ ns [11].

## IV. INVESTIGATION OF THE RF BIAS MODE

Operating the detector using rf bias, should cause a changing operating point over time. Thus, a reduced count rate using an rf current with the same amplitude as a constant dc current is reasonable. To proof that the operating point of the nanowire oscillates with $f_{res}$, we used the detector bias as a reference signal in a synchronous single-photon detection. Figure 6 shows the setup used for our measurements. The biasing and readout of the detector has not been changed in comparison with the previous rf measurements. But now, we use the second channel of the signal generator to modulate the output power of a fast vertical-cavity surface-emitting laser (VCSEL) diode. In doing so, the output light power is proportional to the applied diode current $i_{diode}$. Thus, the RF-SNSPD is biased with a current at frequency $f_{res}$ and the illuminating light is modulated with the frequency $f_{res} + \Delta f$.

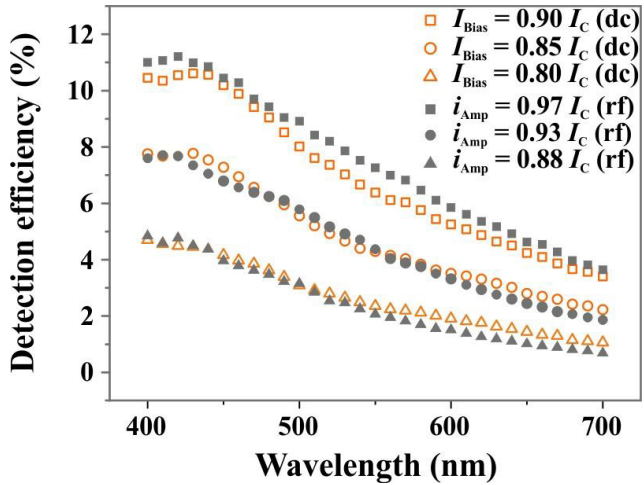

Fig. 5 Comparison of the normalized count rates at different operating points of the detector and different wavelengths of the incident photons. The filled symbols represent measurements using rf bias, the open symbols indicate measurements using dc bias.

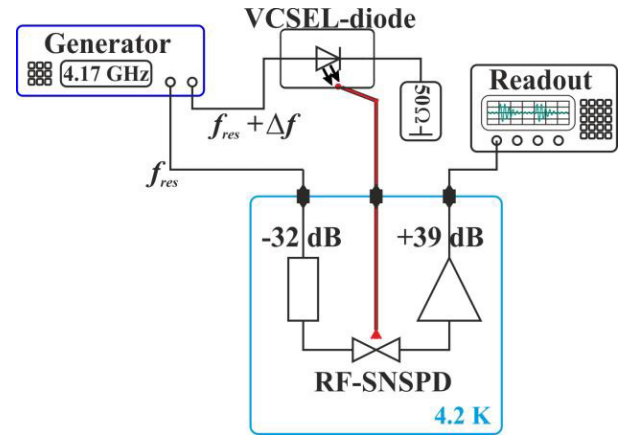

Fig. 6   Schematic of the used setup for the synchronous detection. A signal generator couples the bias signal into the cryostat where it is attenuated by 32 dB at 4.2 K. After transmitted on the CPW, the signal is amplified by 39 dB and recorded with the readout electronics. The second output of the signal generator is used to operate a fast VCSEL diode at a frequency f = $f_{res} + \Delta f$. The modulated light signal of the diode is guided by a multimode fiber to the active area of the detector.

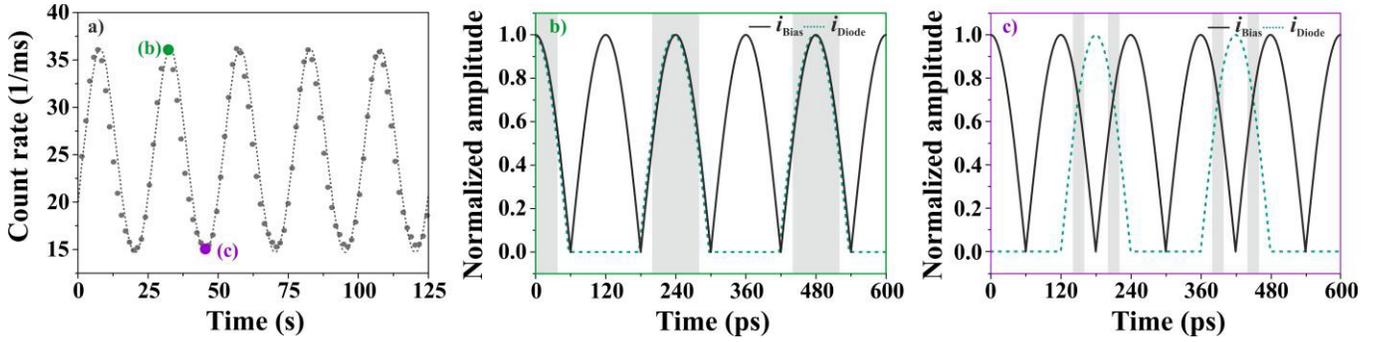

Fig. 7 Results of the synchronous detection of photons. (a) Measured count rates averaged in 10 ms time slots versus time. (b) and (c) Simulation of currents inside the nanowire (solid black line) and current flowing through the diode (dashed lines). Gray shaded are the areas when the detector is photon sensitive. (b) shows both currents in phase, representing the highest count rate. (c) Depicts both currents with 90° phase shift, when the lowest count rate is measured.

For the measurement, we used an RF-SNSPD with the same layout shown in Fig. 1, but with an increased resonant frequency of 4.17 GHz. The difference in frequency between the detector and the light signal was set to $\Delta f = 20$ mHz. Figure 7 (a) shows the measured count rate over time with a gating time of 10 ms for each single measurement. An oscillation of the count rate with a frequency of 40 mHz becomes visible. Thus, we got the doubled frequency of $\Delta f$ as a count-rate modulation. The reason for the frequency doubling is due to the biasing of the nanowire. For the detector it is not important in which direction the current is flowing. So, the operation point of the detector follows the absolute value of the applied bias signal and reaches two times the maximum efficiency during one period. An explanation of the oscillating count rate is given in Fig. 7 (b) and (c).

The maximum count rate is achieved when the bias signal of the nanowire $i_{Bias}$ and the diode current $i_{Diode}$ oscillates in phase, which is simulated in Fig 7 (b). The biasing of the RF-SNSPD follows the absolute function of $f_{res}$. However, the current $i_{Diode}$ can only flow from the anode to the cathode, which is why the negative sine wave is cut off. The regions, when the detector is photon sensitive, are gray shaded. We defined photon sensitivity, when the current through the diode as well as the current in the nanowire reached at least 50 % of the amplitude level.

If the running phase in between the diode and the detector signal reaches 90°, the minimum count rate is measured, which is shown in Fig. 7. ©. The areas when the detector is photon sensitive got smaller, but counting is still possible which is the reason why the count rate in Fig. 7. (a) does not drop down to zero. If the running phase between the detector bias and the light signal is 180°, the count rate reaches the maximum again, which is why the measured frequency of the count-rate modulation in Fig. 7. (a) is doubled in comparison with $\Delta f$.

The successful reconstruction of the amplitude and phase of the modulated light signal, generated by the VCSEL diode, demonstrates the changing operating point of the detector in the rf operation mode. Thus, a reduced count rate of the detector biased with an rf signal of the same amplitude as a stable dc operating point is obvious. On the other hand, the oscillating operating point effectively prevents the detector from latching and enables a very stable operation. Furthermore, the dark-count rate of an array decreases with rising pixel number as the pixels are gated in the GHz-range. Thus, not all pixels are at high bias current at the same time.

## V. Conclusion

We demonstrated a new RF-SNSPD design which enables the operation with dc and rf currents. With this device we showed that the detection efficiency spectrum of the detector is not changed if biased with rf or dc currents. Due to the oscillating operating point in the rf operation mode, a slightly higher amplitude level of the current is needed to reach the same detection efficiency as for the dc operation mode. In a synchronous detection of single photons we further demonstrated the oscillating operating point of the RF-SNSPD with a frequency of two times $f_{res}$.

## Acknowledgment

The authors would like to thank A. Stassen for careful preparation of the samples and also K. Gutbrod for excellent mechanical assistance.

This work was supported in part by the Karlsruhe School of Optics and Photonics (KSOP).

## References

1. Marsili, F and Verma, Varun B and Stern, Jeffrey A and Harrington, S and Lita, Adriana E and Gerrits, Thomas and Vayshenker, Igor and Baek, Burm and Shaw, Matthew D and Mirin, Richard P and others, "Detecting single infrared photons with 93\% system efficiency", Nature Photonics, vol. 7, no. 3, 2013.

2. Rosenberg, D and Kerman, AJ and Molnar, RJ and Dauler, EA, "High-speed and high-efficiency superconducting nanowire single photon detector array", Optics express, vol. 21, no. 2, 2013.
3. Grein, Matthew E and Kerman, Andrew J and Dauler, Eric A and Willis, Matthew M and Romkey, Barry and Molnar, Richard J and Robinson, Bryan S and Murphy, Daniel V and Boroson, Don M, "An optical receiver for the Lunar Laser Communication Demonstration based on photon-counting superconducting nanowires", Advanced Photon Counting Techniques IX, vol. 9492, 2015.
4. Korzh, Boris and Lim, Charles Ci Wen and Houlmann, Raphael and Gisin, Nicolas and Li, Ming Jun and Nolan, Daniel and Sanguinetti, Bruno and Thew, Rob and Zbinden, Hugo, "Provably secure and practical quantum key distribution over 307 km of optical fibre", Nature Photonics, vol. 9, no. 3, 2015.
5. Gemmell, Nathan R and McCarthy, Aongus and Liu, Baochang and Tanner, Michael G and Dorenbos, Sander D and Zwiller, Valery and Patterson, Michael S and Buller, Gerald S and Wilson, Brian C and Hadfield, Robert H, "Singlet oxygen luminescence detection with a fiber-coupled superconducting nanowire single-photon detector", Optics express, vol. 21, no. 4, 2013 .
6. Doerner, Steffen and Kuzmin, Artem and Wuensch, Stefan and Ilin, Konstantin and Siegel, Michael, "Operation of superconducting nanowire single-photon detectors embedded in lumped-element resonant circuits", IEEE Transactions on Applied Superconductivity, vol. 26, no. 3, 2016.
7. Zmuidzinas, J. (2012). Superconducting microresonators: Physics and applications. Annu. Rev. Condens. Matter Phys., 3(1), 169-214.
8. Doerner, Steffen and Kuzmin, Artem and Wuensch, Stefan and Charaev, Ilya and Boes, Florian and Zwick, Thomas and Siegel, Michael, "Frequency-multiplexed bias and readout of a 16-pixel superconducting nanowire single-photon detector array", Applied Physics Letters, vol. 111, no. 3, 2017.
9. Sonnet User's Guide-Manual of the Program Sonnet, Sonnet Software Inc., North Syracuse, NY, USA (2007).
10. J. Yang, A. Kerman, E. Dauler, V. Anant, K. Rosfjord and K. Berggren, "Modeling the Electrical and Thermal Response of Superconducting Nanowire Single-Photon Detectors," Applied Superconductivity, IEEE transactions on applied superconductivity , vol. 17, no. 2, 2007.
11. Kerman, Andrew J and Dauler, Eric A and Keicher, William E and Yang, Joel KW and Berggren, Karl K and Gol'Tsman, G and Voronov, B, "Kinetic-inductance-limited reset time of superconducting nanowire photon counters", Applied physics letters, vol. 88, no. 11, 2006.